\documentclass[12Pt,a4paper]{article}
\usepackage{graphicx}
\usepackage[T1]{fontenc}
\usepackage{mathtools}
\usepackage{amsmath,amssymb,anysize,rotating}
\begin{document}
\title{Generalized inverse xgamma distribution: A non-monotone hazard rate model}
\author{Harsh Tripathi $^{a}$, Abhimanyu Singh Yadav $^{a}$, Mahendra Saha $^{a}$ and Sumit Kumar$^{a}$\footnote{Corresponding author. e-mail: 2016phdsta01@curaj.ac.in}\\
\small $^{a}$Department of Statistics, Central University of Rajasthan, Rajasthan, India}
\date{}
\maketitle
\begin{abstract}
In this article, a generalized inverse xgamma distribution (GIXGD) has been introduced as the generalized version of the inverse xgamma distribution. The proposed model exhibits the pattern of non-monotone hazard rate and belongs to family of positively skewed models. The explicit expressions of some distributional properties, such as, moments, inverse moments, conditional moments, mean deviation, quantile function have been derived. The maximum likelihood estimation procedure has been used to estimate the unknown model parameters as well as survival characteristics of GIXGD. The practical applicability of the proposed model has been illustrated through a survival data of guinea pigs.
\end{abstract}
\section{Introduction}
It is impossible to analysis the reliability characteristics and other properties of any lifetime product without the support of distributions. Lifetime of any item or product must follow a particular distribution shape. In the literature, there are many lifetime distributions are available from inverse family of distributions, viz., inverse exponential [see, keller and kamath ($1982$)], inverse Weibull [see, keller and kamath ($1982$)], inverse Rayleigh [see, Voda ($1972$)], inverse Lindley [see, Sharma et al. ($2014$)], inverse xgamma [see, Yadav et al. ($2018$)] distributions and many more. Also the generalization of inverse family of distributions are available in the literature, such as, generalized inverted exponential [see, Abouammoh et al. ($2009$)], generalized inverse gamma [see, Mead($2013$)], generalized inverse lindley [see Sharma et al. ($2015$)], exponentiated generalized inverse Weibull [see, Elbatal et al. ($2014$)] and many more. Recently, inverse xgamma distribution(IXGD), the inverted version of xgamma distribution [see, Sen et al. ($2016$)] is introduced by Yadav et al. ($2018$). They have also discussed several statistical properties of IXGD and showed the superiority of IXGD among the one parameter inverted family of distributions. If $X$ followed IXGD with scale parameter $\theta$, then the  probability density function (PDF) and cumulative distribution function (CDF) of IXGD are given as [see, Yadav et al. ($2018$)]
\begin{eqnarray}\label{eq1}
f(x;\theta)=\frac{\theta^2}{1+\theta} \frac{1}{x^{2}} {\left(1+\frac{\theta}{2x^{2}}\right)} e^{-\theta/x}~~;x>0,~\theta>0
\end{eqnarray}
\begin{eqnarray}\label{eq2}
F(x)=\left[1+\frac{\theta^2}{2(\theta+1)} \frac{1}{x^{2}}+ \frac{\theta}{\theta+1} \frac{1}{x}\right] e^{-\theta/x}~~;x>0,~\theta>0
\end{eqnarray}

The main aim of this article is to introduced a generalized version of IXGD using the power transformation of IXGD, named as generalized inverse xgamma distribution (GIXGD). After exploration of the literature, we found that no work has been done in the direction to introduced GIXGD. Our aim is to fill up this gap through this present study.\\

Rest of article is organize as follows. In Section $2$, we introduced the PDF and the CDF of GIXGD and also derive the expression of survival and hazard rate functions. Different statistical properties, such as, moments, inverse moments, conditional moments, harmonic mean, mean deviation, quantile function, Bonferroni and Lorenz curves  and a procedure to generate random number from GIXGD have discussed in Section $3$. In Section $4$, estimation of survival and hazard rate functions by using MLEs of the parameters have been discussed. A data set is used to illustrate the applicability of the proposed model in real life scenario in Section $5$. Concluding remarks are made in Section $6$.
 
\section{The model}
The one parameter IXGD is actually the inverted version of XGD [see, Sen et al. ($2016$)], is recently proposed by Yadav et al. ($2018$). They have studied the different statistical properties and estimation of the unknown parameter using different methods of estimation. They have mentioned that IXGD possesses non-monotone hazard rate and also shows the superiority of IXGD among the inverted family of distributions. As we know that, the shape parameter play an important role in flexibility of any lifetime models and hence becomes more realistic for uses in any real life situations. Here, in this article, we have proposed a more flexible model by adding one more parameter $\alpha$, the shape parameter, as the power of IXGD variable.\\ 

If $X$ be a random variable having PDF and CDF mentioned in Equations (1) and (2) respectively, then GIXGD is obtained by using power transformation $Y=X^{1/\alpha}$, where $\alpha$ is the shape parameter. Hence, the PDF and CDF of GIXGD are given as 
\begin{eqnarray}\label{eq3}
f(y; \alpha, \theta)=\frac{\alpha \theta^2}{1+\theta} \frac{1}{y^{(\alpha+1)}} {\left(1+\frac{\theta}{2y^{2\alpha}}\right)} e^{-\theta/y^\alpha};~~~y>0,~\alpha>0,~\theta>0
\end{eqnarray}
and
\begin{eqnarray}\label{eq4}
F(y; \alpha,\theta)=\left[1+\frac{\theta^2}{2(\theta+1)} \frac{1}{y^{(2\alpha)}}+ \frac{\theta}{\theta+1} \frac{1}{y^\alpha}\right] e^{-\theta/y^\alpha};~~~y>0,~\alpha>0,~\theta>0
\end{eqnarray}
respectively. If $\alpha=1$, then GIXGD is converted to IXGD. Reliability characteristics of any life time model is often measured in terms of survival and hazard rate functions. The survival function $S(y;\alpha,\theta)$ and hazard rate function (HRF) $H(y;\alpha,\theta)$ of GIXGD are respectively given below:
\begin{eqnarray}\label{eq5}
S(y;\alpha,\theta)& =& 1-F(y;\alpha,\theta) \nonumber\\
&=&1-\left[1+\frac{\theta^2}{2(\theta+1)} \frac{1}{y^{(2\alpha)}}+ \frac{\theta}{\theta+1} \frac{1}{y^\alpha}\right] e^{-\theta/y^\alpha}
\end{eqnarray}
and
\begin{eqnarray}\label{eq6}
H(y;\alpha,\theta)&=&\frac{f(y;\alpha,\theta)}{S(y;\alpha,\theta)}\nonumber
\\&=&\left[\frac{\frac{\alpha \theta^2}{1+\theta} \frac{1}{y^{(\alpha+1)}} {\left(1+\frac{\theta}{2y^{2\alpha}}\right)} e^{-\theta/y^\alpha}}{1-\left[1+\frac{\theta^2}{2(\theta+1)} \frac{1}{y^{(2\alpha)}}+ \frac{\theta}{\theta+1} \frac{1}{y^\alpha}\right] e^{-\theta/y^\alpha}}\right]
\end{eqnarray}
A typical graphs of PDF and HRF are displayed below for different choices of shape and scale parameters respectively. From the shape of density function, it is clearly observable that GIXGD is positively skewed and uni-modal. Also from the shape of HRF, we observed that initially HRF is increasing and reaches to a peak after that declined slowly, which indicates that the model possesses the non-monotone property of hazard rate. Such behaviour of hazard rate are quite common in reliability studies and clinical trial studies etc.
\begin{figure}[htbp]
\centering
\resizebox{7cm}{6cm}{\includegraphics[trim=.01cm 2cm .01cm .01cm]{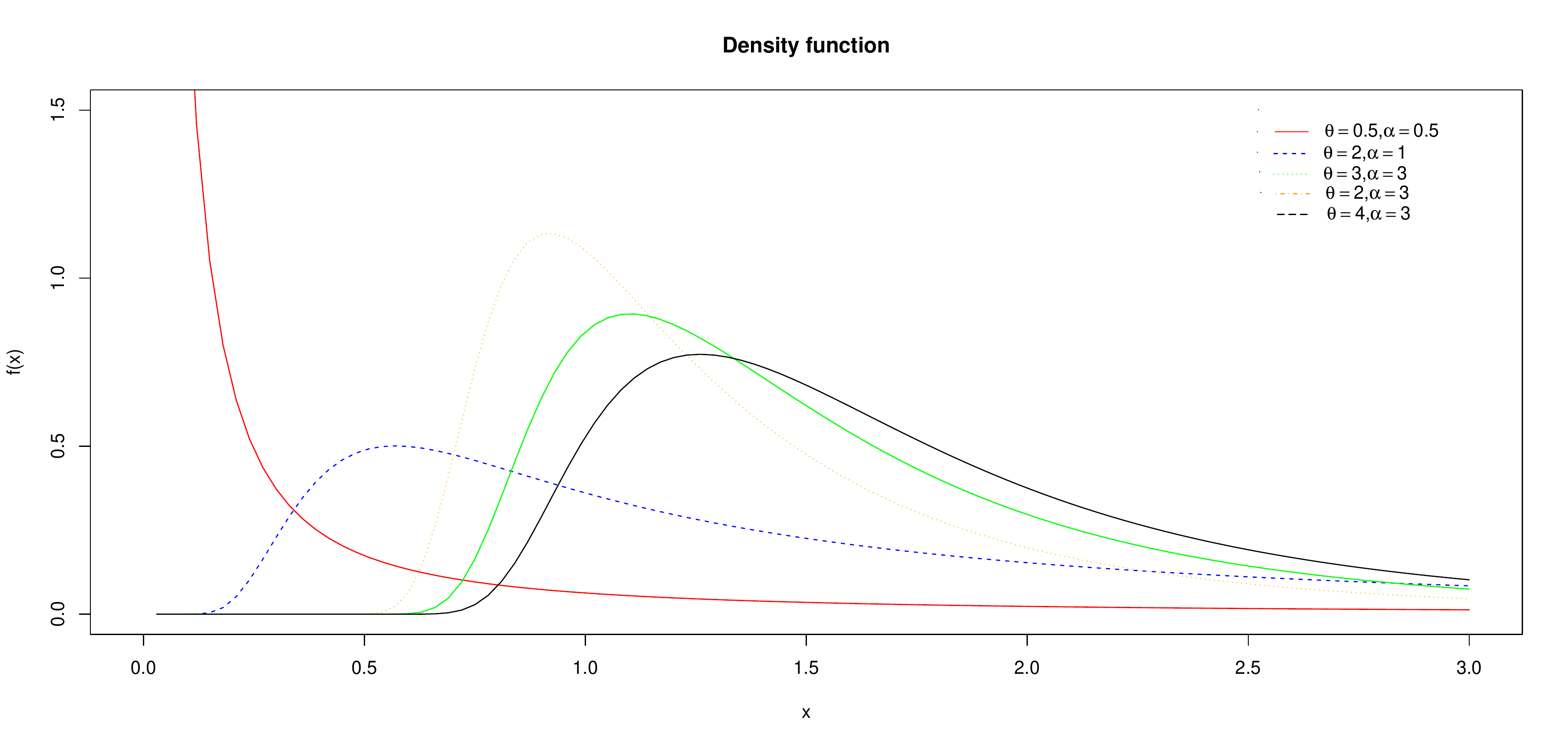}}
\vspace{1cm}
\caption{PDF plot of GIXGD.}
\label{fig1}
\end{figure}
\begin{figure}[htbp]
\centering
\resizebox{7cm}{6cm}{\includegraphics[trim=.01cm 2cm .01cm .01cm]{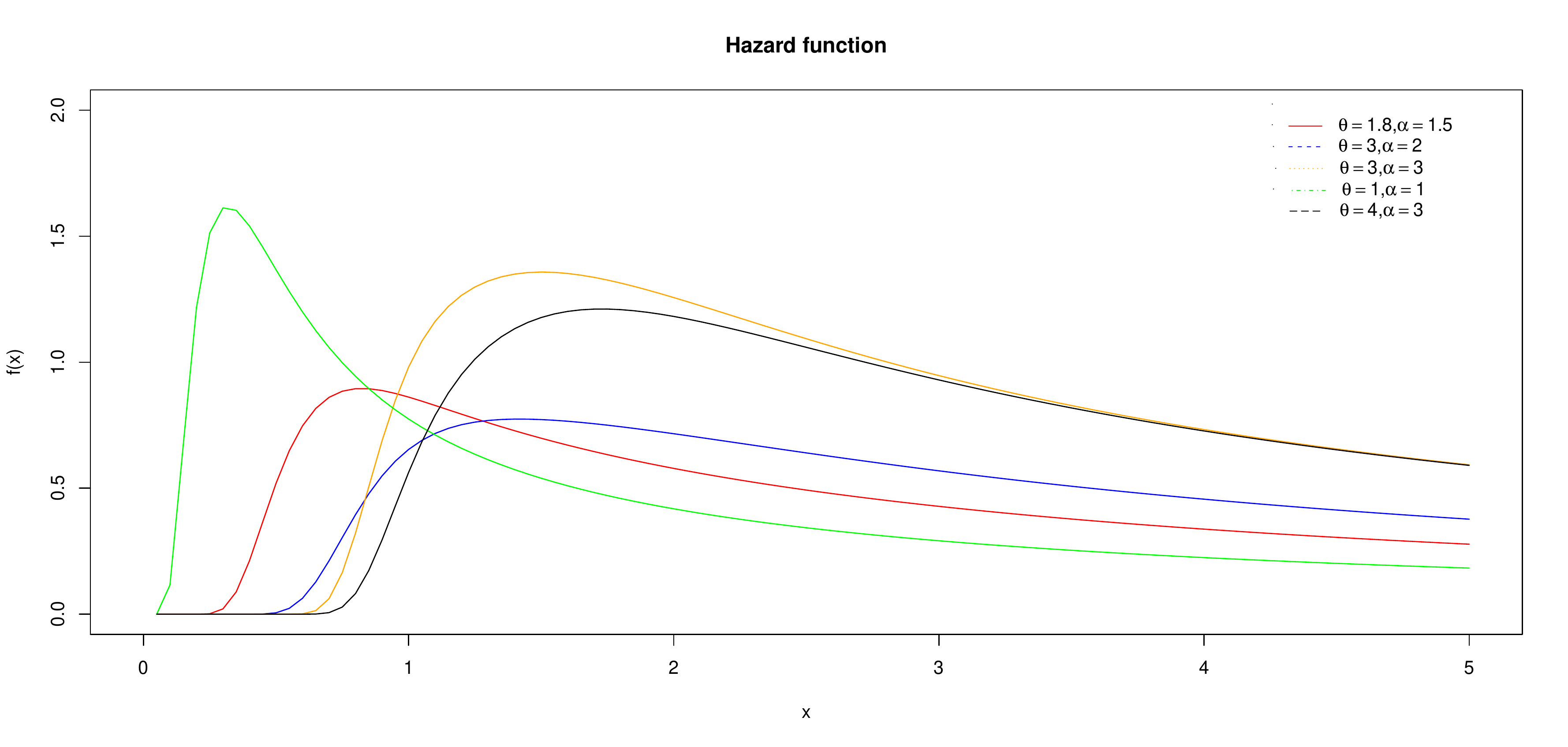}}
\vspace{1cm}
\caption{Hazard plot of GIXGD.}
\label{fig2}
\end{figure}
\section{Some satistical properties of GIXGD}
In this section various properties of GIXGD have obtained viz moments, quantile, conditional moments, mean deviation  and order statistics. Bonferroni and lorenz curves are also calculated.
\subsection{Moments}
Moments are very useful to determine the various properties of a model. Here, we are interested in investing the raw moments of GIXGD. Expression of $c$-th order raw moment is given below
\begin{eqnarray}\label{eq7}
E(y^c)&=&\int\limits_{0}^{\infty}y^c f(y)dy\nonumber
\\&=&\int\limits_{0}^{\infty} y^c \frac{\alpha \theta^2}{1+\theta} \frac{1}{y^{(\alpha+1)}} {\left(1+\frac{\theta}{2y^{2\alpha}}\right)} e^{-\theta/y^\alpha}dy\nonumber
\\&=&\frac{\theta^{(c/\alpha)+1}}{(1+\theta)}\Gamma(1-(c/\alpha))+\frac{1}{2} \frac{\theta^{(c/\alpha)}}{(\theta+1)}\Gamma(3-(c/\alpha))
\end{eqnarray}
$c$-th order raw moment exists iff $\frac{c}{\alpha} \le 1$ and first four central moments can be easily obtained by using the relationship between raw moments and central moments. Hence, Pearson measures of skewness (SK) and kurtosis (K) based on moments can be obtained by using following formulae
$$
\mbox{SK}=\frac{\mu_3^2}{\mu_2^3}~~\mbox{and~~K}= \frac{\mu_4}{\mu_2^2}
$$
where, $\mu_2$, $\mu_3$ and $\mu_4$ are the second, third and fourth central moments respectively.
\subsection{Inverse moments}
The $c$-th order inverse moments about origin of GIXGD is given as
\begin{eqnarray}\label{eq8}
E\left(\frac{1}{y^c}\right)&=&\int\limits_{0}^{\infty}\frac{1}{y^c} f(y) dy\nonumber
\\&=&\int\limits_{0}^{\infty}\frac{1}{y^c} \frac{\alpha \theta^2}{1+\theta} \frac{1}{y^{(\alpha+1)}} {\left(1+\frac{\theta}{2y^{2\alpha}}\right)} e^{-\theta/y^\alpha} dy\nonumber
\\&=&\frac{\theta^2}{2(1+\theta)} \left[\frac{2\Gamma(1+(c/\alpha))}{\theta^{(1+(c/\alpha))}}+\frac{\Gamma(3+(c/\alpha))}{\theta^{(2+(c/\alpha))}}\right]~~~c=1,2,3,...
\end{eqnarray}
The harmonic mean for the random variable is to be obtained from
\begin{eqnarray}\label{eq9}
E\left(\frac{1}{y}\right)&=&\int\limits_{0}^{\infty}\frac{1}{y} f(y) dy
\end{eqnarray}
The above equation can also be calculated from the Equation ($\ref{eq8}$) by putting $c=1$. Hence, after simplification, we get
\begin{eqnarray}\label{eq10}
E\left(\frac{1}{y}\right)=\frac{\theta^2}{2(1+\theta)} \left[\frac{2\Gamma(1+(1/\alpha))}{\theta^{(1+(1/\alpha))}}+\frac{\Gamma(3+(1/\alpha))}{\theta^{(2+(1/\alpha))}}\right]
\end{eqnarray}
\subsection{Conditional moment}
Conditional moments about origin of GIXGD is obtained as
\begin{eqnarray}\label{eq11}
E(Y^n/Y>y)&=&\int\limits_{y}^{\infty} y^n \frac{f(y)}{1-F(y)}\nonumber
\\&=&\frac{1}{1-F(y)}\left[\frac{\theta^{(n/\alpha)+1}}{\theta+1} \gamma{\left(1-\frac{n}{\alpha},\frac{\theta}{y^\alpha}\right)}+\frac{\theta^{(n/\alpha)}}{\theta+1} \gamma{\left(3-\frac{n}{\alpha},\frac{\theta}{y^\alpha}\right)}\right]
\end{eqnarray}
\subsection{Mean deviation}
The mean deviation about mean of random variable $Y$, having density function ($\ref{eq3}$) is obtained as
$$
M.D=\int\limits_{0}^{\infty} |(y-\mu)| f(y) dy 
$$
where, $\mu=E(Y)$. On simplification
\begin{eqnarray}
M.D=2\mu F(\mu)-2\mu+2\int\limits_{\mu}^{\infty}y f(y) dy
\end{eqnarray}
where, $F(\mu)$ stands for CDF of $Y$ up to point $\mu$ and $\int\limits_{\mu}^{\infty}y f(y)dy$ is obtained through use of conditional distribution 
$$
\int\limits_{\mu}^{\infty}y f(y) dy=[1-F(\mu)] E(Y/Y>\mu).
$$
\subsection{Quantile function}
If $Q(p)$ be the quantile of order $p$ of the random variable $X$, then it will be the solution of
\begin{eqnarray}\label{eq12}
F(Q(p))&=&\left[1+\frac{\theta^2}{2(\theta+1)} \frac{1}{Q(p)^{(2\alpha)}}+ \frac{\theta}{\theta+1} \frac{1}{Q(p)^\alpha}\right] e^{-\theta/Q(p)^\alpha}=p
\end{eqnarray}
The degree of long-tail is measured by skewness and while the degree of tail heaviness is measured by kurtosis of the random variable.
 The Bowley measure of skewness [see, Bowley ($1920$)] and Moors measure of kurtosis [see, Moors ($1988$)] based on quantile can be used and are given as 
\begin{eqnarray*}
\mbox{SK}&=&\frac{Q(\frac{3}{4})-2Q(\frac{1}{2})+Q(\frac{1}{4})}{Q(\frac{3}{4})-Q(\frac{1}{4})}
\end{eqnarray*}
\begin{eqnarray*}
\mbox{K}&=&\frac{Q(\frac{7}{8})-Q(\frac{5}{8})+Q(\frac{3}{8})-Q(\frac{1}{8})}{Q(\frac{6}{8})-Q(\frac{2}{8})}
\end{eqnarray*}
\subsection{Bonferroni and Lorenz curves}
Bonferroni and Lorenz curves introduced by Bonferroni. These curves are very helpful in field of income, poverty, reliability, demography and insurance. Let $Y$ be random variable with PDF given Equation ($\ref{eq3}$), then Bonferroni and Lorenz curves are respectively defined as
\begin{eqnarray}\label{eq14}
B(p)&=&\frac{1}{p\mu} \int\limits_{0}^{q} y f(y) dy\nonumber
\\&=&\frac{1}{p\mu}{[\mu-\int\limits_{q}^{\infty}y f(y) dy]}\nonumber 
\\&=&\frac{1}{p\mu}{\left(\mu-\left[\frac{\theta^{(1/\alpha)+1}}{\theta+1} \gamma{\left(1-\frac{1}{\alpha},\frac{\theta}{y^\alpha}\right)}+\frac{\theta^{(1/\alpha)}}{\theta+1} \gamma{\left(3-\frac{1}{\alpha},\frac{\theta}{y^\alpha}\right)}\right]
\right)}
\end{eqnarray}
and
\begin{eqnarray}\label{eq15}
L(p)&=&\frac{1}{\mu} \int\limits_{0}^{q} y f(y) dy\nonumber
\\&=&\frac{1}{\mu}{[\mu-\int\limits_{q}^{\infty} y f(y) dy]}\nonumber
\\&=&\frac{1}{\mu}{\left(\mu-\left[\frac{\theta^{(1/\alpha)+1}}{\theta+1} \gamma{\left(1-\frac{1}{\alpha},\frac{\theta}{y^\alpha}\right)}+\frac{\theta^{(1/\alpha)}}{\theta+1} \gamma{\left(3-\frac{1}{\alpha},\frac{\theta}{y^\alpha}\right)}\right]
\right)}
\end{eqnarray}
and the indices based on these two curves are obtained as
$$B=1-\int\limits_{0}^{1} B(p) dp$$
and
$$G=1-2\int\limits_{0}^{1} L(p)dp$$
respectively, where $B$ and $G$ represents the Bonferroni and Gini indices.
\subsection{Random number generation}
To generate random number from GIXGD ($\alpha,\theta$). The following steps may be used.
\begin{enumerate}
\item Generate $U_i$ from uniform(0,1) distribution $(i=1,2,3...,n)$.\\
\item Generate $V_i$ from $gamma(1,\theta)$ distribution $(i=1,2,3...,n)$.\\
\item Generate $W_i$ from $gamma(3,\theta)$ distribution $(i=1,2,3...,n)$.\\
\item If $U_i\leq\frac{\theta}{\theta+1}$, set $Z_i=V_i$, otherwise set $Z_i=W_i$.\\
\item $X=(1/Z)$ be the random numbers from IXGD.
\item $Y=X^{1/\alpha}$ be random numbers GIXGD.
\end{enumerate}
If we take $\alpha=1$, then the algorithm of generating random number from GIXGD is  same as that of IXGD.
\section{Estimation of survival and hazard rate functions of GIXGD}
In this section, we have used maximum likelihood estimation (MLE) to estimate the unknown parameters as well as survival function $S(y)$ and hazard rate function $H(y)$ for the GIXGD($\alpha,\theta$). Let $Y_{1},~Y_{2}, \cdots,~Y_{n}$ be a random sample of size $n$ from Equation (\ref{eq3}). Then, the likelihood function for the observed random sample $y_{1},~y_{2}, \cdots,~y_{n}$ is given as
\begin{eqnarray}
L(\alpha,\theta|y)=\prod\limits_{i=1}^{n}\left[\frac{\alpha \theta^2}{1+\theta} \frac{1}{y_i^{(\alpha+1)}} {\left(1+\frac{\theta}{2y_i^{2\alpha}}\right)} e^{-\theta/y_i^\alpha}\right]
\end{eqnarray}
Taking logarithm on both sides of Equation ($17$), we have
\begin{eqnarray}
\log L(\alpha,\theta|y)=n\log(\alpha)+2n \log{\theta}-n \log{(\theta+1)}+\sum\limits_{i=1}^{n} \log(\frac{1}{y_i^{(\alpha+1)}})+\sum\limits_{i=1}^{n} \log{(1+\frac{\theta}{2 y_i^{2\alpha}})}-\sum\limits_{i=1}^{n}\frac{\theta}{y_i^\alpha}
\end{eqnarray}
Partial derivatives of the log-likelihood function with respect to $\alpha$ and $\theta$ and equating to zero yield the estimate of $\alpha$ and $\theta$ respectively, i.e.,
\begin{eqnarray}
\frac{\partial\log L(\alpha,\theta|y)}{\partial\alpha}=\frac{n}{\alpha}-\sum\limits_{i=1}^{n} \log y_i-\sum\limits_{i=1}^{n}\frac{\theta y^{-2\alpha} \log y}{\left(1+\frac{\theta}{2 y_i^{2\alpha}}\right)}+\theta\sum\limits_{i=1}^{n} y_i^{-\alpha} \log y_i=0
\end{eqnarray}
\begin{eqnarray}
\frac{\partial\log L(\alpha,\theta|y)}{\partial\theta}=\frac{2n}{\theta}-\sum\limits_{i=1}^{n}\frac{1}{y_i^\alpha}+\sum\limits_{i=1}^{n}\frac{(1/2y_i^{2\alpha})}{\left(1+\frac{\theta}{2 y_i^{2\alpha}}\right)}-\frac{n}{\theta+1}=0
\end{eqnarray}
Equating these partial derivatives to zero did not yield closed-form solutions for the MLEs and thus a numerical method is used for solving these equations simultaneously. Substituting the MLEs ($\hat{\alpha}_{mle},\hat{\theta}_{mle}$) of ($\alpha,\theta$) and using the invariance properties of MLES, we can get the estimators of $S(y)$ and $H(y)$ as
\begin{eqnarray}
\hat{S}(y)_{mle}=1-\left[1+\frac{\hat{\theta}_{mle}^2}{2(\hat{\theta}_{mle}+1)} \frac{1}{y^{(2\hat{\alpha}_{mle})}}+ \frac{\hat{\theta}_{mle}}{\hat{\theta}_{mle}+1} \frac{1}{y^{\hat{\alpha}_{mle}}}\right] e^{-\hat{\theta}_{mle}/y^{\hat{\alpha}_{mle}}}
\end{eqnarray}
and
\begin{eqnarray}
\hat{H}(y)_{mle}=\left[\frac{\frac{\hat{\alpha}_{mle} \hat{\theta}_{mle}^2}{1+\hat{\theta}_{mle}} \frac{1}{y^{(\hat{\alpha}_{mle}+1)}} {\left(1+\frac{\hat{\theta}_{mle}}{2y^{2\hat{\alpha}_{mle}}}\right)} e^{-\hat{\theta}_{mle}/y^{\hat{\alpha}_{mle}}}}{1-\left[1+\frac{\hat{\theta}_{mle}^2}{2(\hat{\theta}_{mle}+1)} \frac{1}{y^{(2\hat{\alpha}_{mle})}}+ \frac{\hat{\theta}_{mle}}{\hat{\theta}_{mle}+1} \frac{1}{y^{\hat{\alpha}_{mle}}}\right] e^{-\hat{\theta}_{mle}/y^{\hat{\alpha}_{mle}}}}\right]
\end{eqnarray}
respectively.
\section{Application}
Here, we have considered one data set, initially considered by Bjerkedal ($1960$) which represents the survival times (in days) guinea pigs with different doses of tubercle bacilli. The regimen is common logarithmic of number of bacillary units per $0.5$ $ml.$ Corresponding to $6.6$ regimen, there were $72$ observations given below.
$$12,15,22,24,24,32,32,33,34,38,38,43,44,48,52,53,54,54,55,56,57,58,58,59,60,60$$
$$60,60,61,62,63,65,65,67,68,70,70,72,73,75,76,76,81,83,84,85,87,91,95,96,98,99$$
$$109,110,121,127,129,131,143,146,146,175,175,211,233,258,258,263,297,341,341,376$$
At first we have checked whether the considered data set is actually comes from GIXGD or not by goodness-of-fit test and compared the fit with the following lifetime distributions:
\begin{itemize}
    \item Generalized inverse xgamma Distribution (GIXGD)
    \item Inverse Lindley distribution (ILD)
	\item Inverse xgamma distribution (IXGD)
	\item Inverse Weibull distribution(IWD)
	\item Inverted exponential distribution (IED)
	\item Generalized exponential distribution (GED)
	\item Gamma distribution (GD)
	
\end{itemize}
This procedure is based on the Kolmogorov-Smirnov (K-S) statistic and it compares an empirical and a theoretical model by computing the maximum absolute difference between the empirical and theoretical CDFs. Note that, K-S statistic to be used only to verify the goodness-of-fit and not as a discrimination criteria. Therefore, we consider four discrimination criteria based on the log-likelihood function evaluated at the maximum likelihood estimates of the parameters. The criteria are: Akaike information criterion (AIC), corrected Akaike information criterion (CAIC), Hannan-Quinn information criterion (HQIC), Bayesian information criterion (BIC). The model with least AIC, CAIC, HQIC, BIC and KS is treated as best model. The obtained measures are reported in Table $1$ which indicates that the GIXGD is best choices among one parameter as well as two parameters family of distributions. Hence, GIXGD may be chosen as an alternative model. Again the MLEs of ($\alpha,\theta$) and using invariance property $S(y)$ and $H(y)$ are obtained for any specified value of $y$, say,$y=54,~70,~99,~112$ are obtained from the above data set, displayed in Table $2$. 
\begin{table}[htbp]
	\centering
	\caption{\bf The model fitting summary for the considered data set.}
	\begin{tabular}{cccccccc}\\
		\hline
		Model & MLE   &-LogL  & AIC   & BIC  & HQIC  & CAIC   & K-S \\
		\hline
		GIXGD    &[1.624156, 641.7531] & 391.9910 & 787.9820 & 792.5353 & 789.7947 & 793.5353 &0.143219 \\
		ILD    & 61.06575 & 402.6685 & 807.3371 & 809.6137 & 808.2434 & 810.6137 & 0.184594 \\
		IXGD    & 61.844 & 402.8761 & 807.7522 & 810.0289 & 808.6585 & 811.0289 & 0.187181 \\
		IWD  & [1.414755, 283.831] & 420.1391 & 844.2782 & 848.8316 & 846.0909 & 849.8316 & 0.138098 \\
		IED    & 0.01663913 & 402.6718 & 807.3437 & 809.6203 & 808.2500 & 810.6203 & 0.184658 \\
		GED    & [2.4741, 58.95436] & 393.1103 & 790.2205 & 794.7739 & 792.0332 & 795.7739 & 0.132821 \\
		GD  & [2.081276, 0.02084991] & 394.2476 & 792.4952 & 797.0485 & 794.3079 & 798.0485 & 0.998295 \\
	\hline
	\end{tabular}%
	\label{tab:addlabel}%
\end{table}%
\begin{table}[htpp]
\centering
\caption{\bf Data analysis using MLEs of the parameters}
\begin{tabular}{|l|l|l|l|l|}\hline
\multicolumn{1}{|c|}{}&
\multicolumn{2}{|c|}{MLEs of the parameters}&
\multicolumn{2}{|c|}{MLEs of $S(y;\alpha,\theta)$ and $H(y;\alpha,\theta)$}\\
\cline{2-5}
$Y=y$   & $\hat{\alpha}$ & $\hat{\theta}$  & $\hat{S}(y;\alpha,\theta)$ & $\hat{H}(y;\alpha,\theta)$ \\
\hline
$y=54$  & $1.624157$     & $641.7557$      & $0.625921$                 & $0.017661$ \\
\hline
$y=70$  &$1.624157$      & $641.7557$      & $0.475482$                 & $0.016507$ \\
\hline
$y=99$  &$1.624157$      & $641.7557$      & $0.307587$                 & $0.013570 $ \\
\hline
$y=112$ &$1.624157$      & $641.7557$      & $0.259806$                 & $0.012426$ \\
\hline
\end{tabular}%
\end{table}
All the computations has done by using $R$ software[see, Ihaka and Gentleman ($1996$)].
\section{Concluding remarks}
In this article, we have proposed a new positively skewed probability distribution, namely, GIXGD by considering the power transformation of IXGD, introduced by Yadav et al. ($2018$). Several statistical properties have been derived. MLEs of the unknown parameters as well as survival and hazard rate functions using invariance property of MLE have been obtained for the proposed model GIXGD. Also the estimation of survival and hazard rate functions using different methods of estimation, viz., ordinary and weighted least square estimation, Cram\`er-von-Mises estimation and maximum product of spacing estimation and the Monte Carlo simulation study in order to compare the performance among the estimators in mean squared error sense are in process.  Finally, a real data set has been analyzed for illustration purposes of the proposed study. Estimation of the parameters and the reliability characteristics may be further studied under different types of censoring scheme. Also the Bayesian estimation may further be considered with suitable priors and loss functions in future.\\
\section{References}
\begin{enumerate}
\item Abouammoh, A. M., Alshingiti, A. M. (2009). Reliability estimation of generalized inverted exponential distribution. {\it Journal of Statistical Computation and Simulation}, {\bf 79(11)}, 1301-1315.
\item Bowley A. L. (1920). Element of Statistics. P.S. King and Son, Ltd., New York.
\item Bjerkedal, T. (2009). Acquisition of resistance in guinea pigs infected with different doses of virulent tubercle bacilli. {\it American journal of Hygiene}, {\bf Vol.72)}, 130-148.
\item Elabatal, I., Muhammed, Hiba Z.(2014). Exponentiated generalized inverse weibull distribution.{\it Applied mathematical sciences}{\bf Vol-8(21)}, 3997-4012
\item Ihaka, R. and Gentleman, R. (1996). R: A language for data analysis and graphics. {\it Journal of Computational and Graphical Statistics}, {\bf 5}, 299-314.
\item Keller, AZ and Kamath, AR(1882). Reliability analysis of CNC machine tools. {\it Reliability engineering} {\bf ,3}, 449-473
\item Mead, M. E. (2015). Generalized inverse gamma distribution and its application in reliability. {\it Communications in Statistics-Theory and Methods}, {\bf 44(7)}, 1426-1435.
\item Moors, J. J. (1988). A quantile alternative for kurtosis, {\it Journal of Royal Statistical Society Series D}, {\bf 37}, 25-32.
\item Sen, S., Maiti, S. S. and Chandra, N. (2016). The xgamma Distribution: statistical properties and Application {\it Journal of Modern Applied Statistical Methods}, {\bf 15(1)}, 774-788.
\item Sharma, V. K., Singh, S. K., Singh, U., Agiwal, V. (2014). The inverse Lindley distribution: a stress-strength reliability model with application to head and neck cancer data.{\it Journal of Industrial and Production Engineering}, {\bf 32(3)}, 162-173.
\item Sharma, V. K., Singh, S. K., Singh, U., Merovci, F. (2015). The generalized inverse Lindley distribution: A new inverse statistical model for the study of upside-down bathtub data.{\it Communications in Statistics-Theory and Methods}, {\bf 45(19)}, 5709-5729.
\item Voda, V.G.(1972). On the inverse Rayleigh random variable. {\it Rep. Stat. Appl. Res. Jues,}{\bf 19(4)}, 13-21
\item Yadav, A. S., Maiti, S. S. and Saha, M. (2018). The inverse xgamma distribution: statistical properties and different methods of estimation. {\it Austrian Journal of Statistics}, communicated.
\end{enumerate}
\end{document}